\documentclass[a4paper]{article}

\usepackage{INTERSPEECH2021}
\usepackage{cite}

\title{Mixture of orthogonal sequences made from extended time-stretched pulses enables measurement of involuntary voice fundamental frequency response to pitch perturbation}
\name{Hideki Kawahara$^1$, Toshie Matsui$^2$, Kohei Yatabe$^3$, Ken-Ichi Sakakibara$^6$, 
Minoru Tsuzaki$^4$, \\ Masanori Morise$^5$, and Toshio Irino$^1$}
\address{
  $^1$Wakayama University, Wakayama, 640-8510 Japan\\
  $^2$Toyohashi University of Technology, Aichi, 441-8580 Japan\\
  $^3$Waseda University, Tokyo, 169-8555 Japan\\
  $^6$Health Science University of Hokkaido, Hokkaido, 061-0293 Japan\\
  $^4$Kyoto City University of Arts, Kyoto 610-1197 Japan\\
  $^5$Meiji University, Tokyo, 164-8525 Japan}
\email{kawahara@wakayama-u.ac.jp, tmatsui@cs.tut.ac.jp, k.yatabe@asagi.waseda.jp,
kis@hoku-iryo-u.ac.jp,minoru.tsuzaki@kcua.ac.jp, mmorise@meiji.ac.jp, 
irino@wakayama-u.ac.jp}

\begin{document}

\maketitle
\begin{abstract}
Auditory feedback plays an essential role in the regulation of the fundamental frequency of voiced sounds.  The fundamental frequency also responds to auditory stimulation other than the speaker's voice.  We propose to use this response of the fundamental frequency of sustained vowels to frequency-modulated test signals for investigating involuntary control of voice pitch. This involuntary response is difficult to identify and isolate by the conventional paradigm, which uses step-shaped pitch perturbation.  We recently developed a versatile measurement method using a mixture of orthogonal sequences made from a set of extended time-stretched pulses (TSP).  In this article, we extended our approach and designed a set of test signals using the mixture to modulate the fundamental frequency of artificial signals. For testing the response, the experimenter presents the modulated signal aurally while the subject is voicing sustained vowels. We developed a tool for conducting this test quickly and interactively. We make the tool available as an open-source and also provide executable GUI-based applications. Preliminary tests revealed that the proposed method consistently provides compensatory responses with about 100~ms latency, representing involuntary control. Finally, we discuss future applications of the proposed method for objective and non-invasive auditory response measurements. 
\end{abstract}

\section{Introduction}
The fundamental frequency 
($f_\mathrm{o}$)\footnote{We use the symbol $f_\mathrm{o}$ to represent
the fundamental frequency adopting the discussion in the forum article
\cite{titze2015jasaforum}.} of sustained vowels respond to frequency modulation of aurally presented sounds\cite{larson2005Jasa,BEHROOZMAND201289,PATEL2016772.e33}.
This response, combined with our recently developed system analysis
method\cite{kawahara2020simultaneous,kawahara2021icassp}
provides a versatile new tool for investigating the auditory-to-speech chain.
Preliminary tests illustrated that the proposed method provides accurate measurement of the involuntary response to $f_\mathrm{o}$ perturbation of the aurally presented test signals.
The measured latency of the response was around 100~ms and compensatory to the perturbation.
The goal of this article is to introduce our method and to demonstrate the feasibility of the proposed method. 
We developed an easy-to-use tool and made it available as an open-source for readers to be able to replicate and verify our results.

\subsection{Background and related work}

Without proper regulation, we are not able to keep the fundamental frequency
of the voice (for example, sustained vowels) constant\cite{titze1994book}.
Auditory feedback plays an essential role in 
this regulation\cite{jones2008auditory,TOURVILLE20081429,houde2013PNAS}.
Vibrato, which makes singing voice attractive, also involves auditory feedback in production\cite{leydon2003role,titze2002reflex}.
Despite decades of research on voice fundamental frequency control mechanisms, it still is a hot topic\cite{larson2016sensory,behroozmand2020modulation,murray2020relationships,peng2021causal}.
Note that the target of the regulation is not the $f_\mathrm{o}$ value.
The target is the perceived pitch and is a psychological attribute,\cite{Moore2013book}.
For periodic signals, $f_\mathrm{o}$ value is the perceived pitch's physical correlate.
In other words, we can observe the perceptual attribute, pitch, directly using the $f_\mathrm{o}$ value
of the produced voice.
The regulation of voice pitch consists of voluntary and involuntary control\cite{hain2000instructing,zarate2010neural}.
The shifted pitch paradigm\cite{burnett1997voice} used in these studies has difficulty investigating this involuntary response.

The first author proposed to use a pseudo-random signal\cite{schroeder1979integrated} to perturb the
$f_\mathrm{o}$ of the fed-back voice.
It enabled to make the test signal unpredictable and to derive
the impulse response of the auditory-to-voice $f_\mathrm{o}$ chain\cite{kawahara1994interactions,kawahara1996icslp}.
This unpredictability enabled measurement of involuntary response to pitch perturbation.
However, it was difficult for others to replicate the test
because it required a complex combination of hardware and
software tools.
The procedure also consisted of several drawbacks due to
available technology in the 1990s.
For example, we measured the response to pitch perturbation using the maximum length sequence (MLS)\cite{schroeder1979integrated}.
Selection of MLS among other TSP signals~\cite{aoshima1981jasa,dunn1993distortion,farina2000simultaneous,stan2002comparison,guidorzi2015impulse} was inevitable to
make the test signal unpredictable.
However, MLS has difficulty in measuring systems with non-linearity\cite{farina2000simultaneous,stan2002comparison}.
Conventional pitch extractors are the other source of problem.
They introduced non-linear and unpredictable distortions in the
extracted $f_\mathrm{o}$ trajectories.
\subsection{Contribution of the current work}
We succeeded in making test signals which are unpredictable and do not have MLS's difficulty.
Our new system analysis method uses a new extended TSP called CAPRICEP (Cascaded All-Pass filters with RandomIzed CEnter frequencies and Phase Polarities)\cite{kawahara2021icassp}.
We used CAPRICEP and developed an auditory-to-speech chain analysis system by adopting the simultaneous measurement method of linear, non-linear, and random responses\cite{kawahara2020simultaneous}.
We developed an instantaneous frequency-based $f_\mathrm{o}$ analysis method instead of using conventional pitch extractors and removed the above-mentioned distortions.
The combination of these analysis methods and substantially advanced
computational power removed all the difficulties in measuring the auditory-to-speech chain response and resulted in an easy-to-use tool for conducting experiments. 
The tool is open-sourced and available from the first author's GitHub repository\cite{kawahara2020gitHk}.

The following section introduced the proposed method with illustrative plots of component procedures.
Then, we introduce a GUI-based application for conducting experiments based on the proposed method quickly and interactively.
The section shows preliminary test examples to illustrate how to use the tool and how to analyze the results.
Finally, we discuss the further application of this method for objective and non-invasive auditory response measurements.

The associated media provides a movie showing how an interactive test tool for conducting the proposed method works to readers.
The media also consists of an example recording of a test session and link to the tool for readers to investigate details and to verify our results.

\section{Method}
\begin{figure}[tbp]
    \centering
    \includegraphics[width=0.9\hsize]{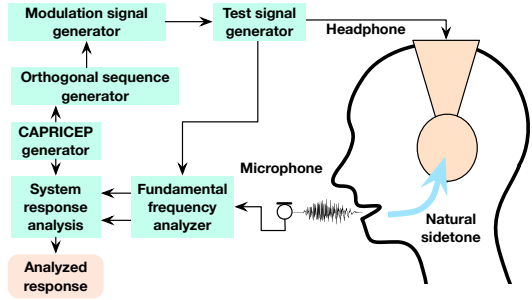}\\
    \vspace{-3mm}
    \caption{Schematic diagram of the involuntary response measurement to pitch perturbation using the proposed method.}
    \label{fig:experimentSetting}
    \vspace{-2mm}
\end{figure}
Figure~\ref{fig:experimentSetting} shows a schematic diagram of the experimental setting of the proposed method.
The task of the subject is to keep voicing a sustained vowel at a constant pitch while exposed to the test sound using a headphone.
The analysis of the response to pitch perturbation uses $f_\mathrm{o}$ values of both the test and the voiced sounds.
A special design enabled analysis of the involuntary response to pitch perturbation.

\subsection{Test signal design}
Test signal design consists of the following four steps.
The first step generates extended TSP signals (unit-TSPs) based CAPRICEP\cite{kawahara2021icassp}.
The ``\textbf{CAPRICEP generator}'' in Fig.~\ref{fig:experimentSetting} does this process.
The second step periodically ($t_\mathrm{r}$ represents the period) allocates unit-TSPs using a set of orthogonal series to yield a set of (after post-processing) orthogonal sequences\cite{kawahara2020simultaneous}.
The ``\textbf{Orthogonal sequence generator}'' in Fig.~\ref{fig:experimentSetting} does this process.
The third step mixes and smooths the orthogonal sequences to make a modulation signal for frequency modulation.
The ``\textbf{Modulation signal generator}'' in Fig.~\ref{fig:experimentSetting} does this process.
The fourth step frequency modulates the carrier signals such as a single sine wave and signals consisting of harmonically related multiple sine waves.
The ``\textbf{Test signal generator}'' in Fig.~\ref{fig:experimentSetting} does this process.

The third step generates a modulation signal having the fundamental period 
$4\!\!\times\!\! t_\mathrm{r}$ with and without smoothing.
Filtering using time-reversed unit-TSPs followed by post-processing using the set of orthogonal series recovers periodic pulse sequences with the period $t_\mathrm{r}$ and a pulse sequence with the period $4\!\! \times \!\!t_\mathrm{r}$.

\subsection{Analysis of response}
The fundamental frequency analyzer uses analytic signals which are tuned to the target fundamental frequency. 
The instantaneous frequency of the filtered signal is the temporally varying fundamental frequency.
We used a six-term cosine series for the envelope of the analytic signals\cite{kawahara2017interspeechGS} for calculating clean instantaneous frequency.
The ``\textbf{Fundamental frequency analyzer}'' in Fig.~\ref{fig:experimentSetting} does this process.

This process applies the pulse recovery and correlation-cancellation procedures described in the previous section.
It yields the perturbation pulse shape from the electrically fed-back signal.
It also yields the response to the perturbation from the recorded sound.
The ``\textbf{System response analysis}'' in Fig.~\ref{fig:experimentSetting} does this process.

\subsection{Design and analysis parameters}
We designed the temporal distribution of the power of unit-TSP to have the raised cosine shape\cite{kawahara2021icassp}.
The nominal duration of the unit-TSP was 400~ms.
We set the allocation interval of unit-TSPs as 16384 samples (371.5~ms for 44100Hz sampling).
We selected three items from the unit-TSP pool to generate one set of orthogonal sequences.
Note that the mixtures of three sequences made from the different sets of unit-TSPs are independent.
We set the total duration of the test signal 20~s.
This setting provides about 60 repetitions of measurement for calculating one impulse response.
This repetition reduces the observation error by about 1/8 in terms of standard deviation.

\subsection{Generated signals and analyzed signals}
This section uses the generated signals in each procedure shown in Fig.~\ref{fig:experimentSetting} to illustrate its function.
\begin{figure}[tbp]
    \centering
    \includegraphics[width=0.9\hsize]{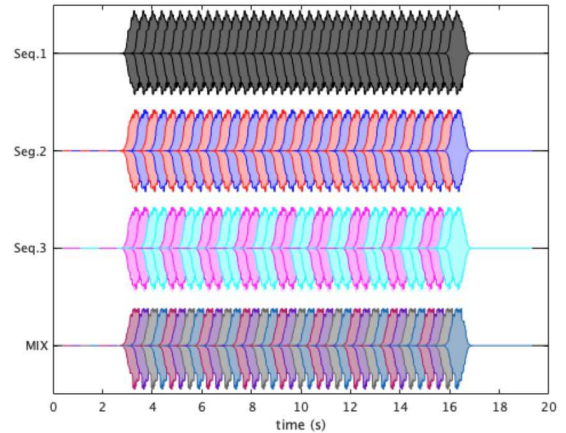}\\
    \vspace{-2mm}
    \caption{Periodic allocation of extended unit-TSPs and making of the mixed signal. Elements are overlap and added.}
    \label{fig:mixedSignal}
    \vspace{-2mm}
\end{figure}
Figure~\ref{fig:mixedSignal} illustrates the function of ``\textbf{Orthogonal sequence generator}.''
The labels ``\textbf{Seq.1},'' ``\textbf{Seq.2},'' and ``\textbf{Seq.3}'' are orthogonal sequences made from three different unit-TSPs.
For ``\textbf{Seq.1},'' we allocated the first unit-TSP periodically with the same polarity.
For ``\textbf{Seq.2},'' we allocated the second unit-TSP periodically inverting the polarity each time.
For ``\textbf{Seq.3},'' we allocated the third unit-TSP periodically inverting the polarity every other time.
We overlap and added each allocation to yield the sequence.
We mixed all sequences to make the signal ``\textbf{MIX}'' for the following process.

\begin{figure}[tbp]
    \centering
    \includegraphics[width=0.9\hsize]{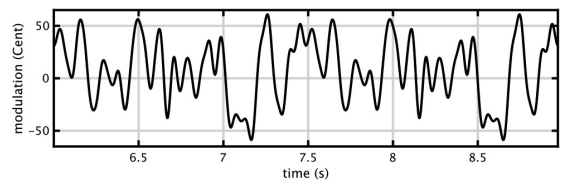} \\
    \vspace{-2mm}
    \caption{Smoothed modulation signal prepared for frequency modulation.}
    \label{fig:modulationSignal}
    \vspace{-2mm}
\end{figure}
Figure~\ref{fig:modulationSignal} shows an example of the generated signal by ``\textbf{Modulation signal generator}.'' 
We smoothed the mixed signal (``\textbf{MIX}'' in Fig.~\ref{fig:mixedSignal}) generated by the preceding procedure using the six-term cosine series\cite{kawahara2017interspeechGS}.  
This smoother provides more than 114~dB suppression of interfering signals outside of the main lobe of the frequency response.
The plot shows a portion of the signal with the length of eight allocation periods.
We set the standard deviation of the modulation signal to 25~Cent.
This setting is to design the speed of $f_\mathrm{o}$ transition of this modulation signal not to exceed that observed in natural speech sounds\cite{xu2002maximum}.
Note that this smoothed signal modulates the $f_\mathrm{o}$ represented in log-frequency because a set of linear differential equations approximate the $f_\mathrm{o}$ dynamics well when using the log-frequency representation\cite{fujisaki1988note}.

\begin{figure}[tbp]
    \centering
    \includegraphics[width=0.9\hsize]{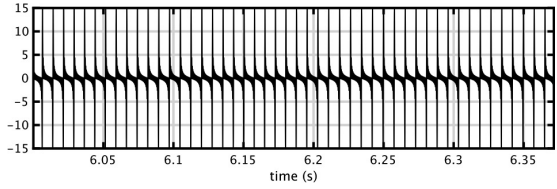}\\
    \vspace{-2mm}
    \caption{Frequency modulated test signal with multiple harmonic components. (Sine phase)}
    \label{fig:testSinesSignal}
    \vspace{-2mm}
\end{figure}
Figure~\ref{fig:testSinesSignal} shows an example of the test signal consisting of twenty harmonic components generated using the ``sine'' phase.
This is the output of ``\textbf{Test signal generator}'' in Figure~\ref{fig:experimentSetting}.
The plot shows the portion with one allocation interval's width.
Note that the waveform deformation caused by the frequency modulation is not visible in this plot because the magnitude of the modulation is less than 7\% of the fundamental period.
The headphone converts this test signal to the test sound and presents the sound to the subject.

\begin{figure}[tbp]
    \centering
    \includegraphics[width=0.442\hsize]{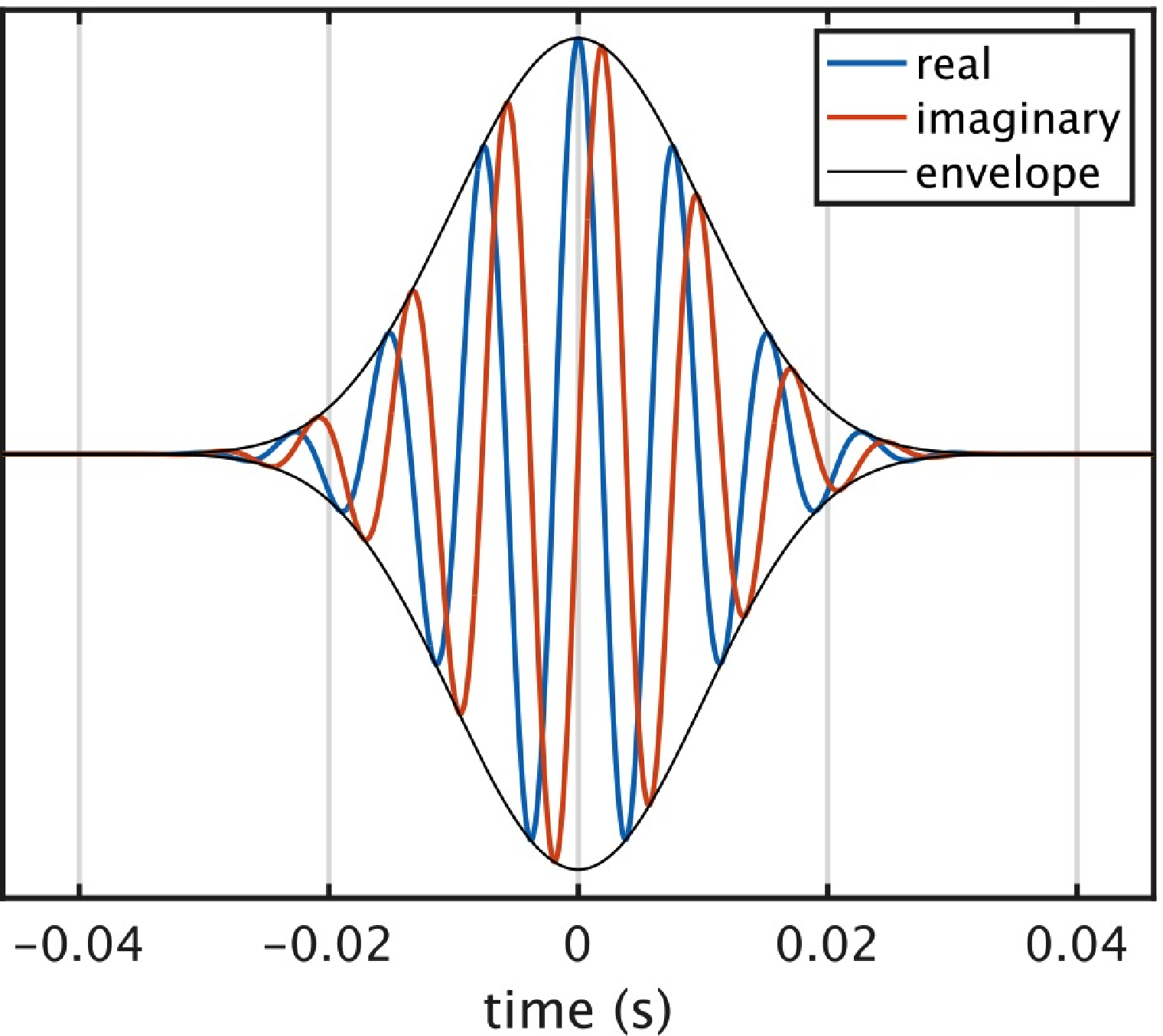}
    \hfill
    \includegraphics[width=0.518\hsize]{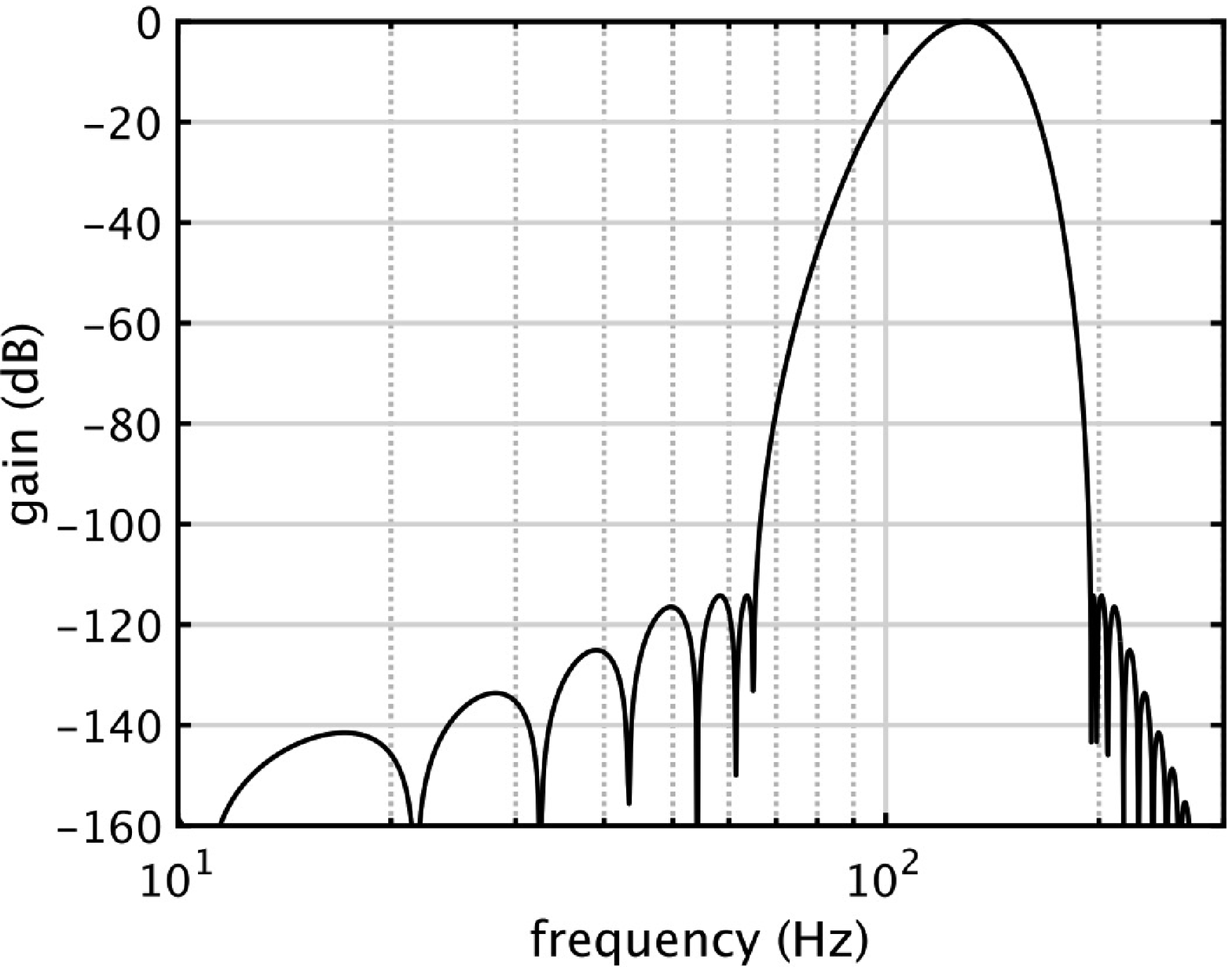}\\
    \vspace{-2mm}
    \caption{Impulse response (left) and gain (right) of the analytic signal filter for analyzing 130~Hz sinusoidal component.}
    \label{fig:filters}
    \vspace{-2mm}
\end{figure}
Figure~\ref{fig:filters} shows the impulse response $h_c[n]$ and the frequency gain response of the filter for $f_\mathrm{o}$ analysis.
The procedure ``\textbf{Fundamental frequency analyzer}'' uses this filter.
The following equation provides the instantaneous frequency $f_i[n]$ of the filtered output $y[n]$, where $n$ represents the index of the discrete-time signal.
\begin{align}\label{eq:instFq}
   f_i[n] & = \angle\left[\frac{y[n + 1]}{y[n]} \right] 
   \cdot \frac{f_s}{2\pi} 
\end{align}
where $\angle[a]$ represents the argument of a complex number $a$ and $f_s$ represents the sampling frequency of the discrete-time signal.
Because the fundamental frequency of the test signal and the sustained vowel are known in advance, we avoid using conventional pitch extractors.
Those extractors consist of pre and post-processing procedures and introduce non-linear and unpredictable distortions.
Equation~\ref{eq:instFq} does not introduce such distortions and yields the $f_\mathrm{o}$ values at the audio sampling rate.
The impulse response $h_c[n]$ design is crucial for Eq.~\ref{eq:instFq} to yield accurate instantaneous frequency values\cite{kawahara2017pitfall}.
We used the six-term cosine series\cite{kawahara2017interspeechGS} to design the envelope of $h_c[n]$.

\begin{figure}[tbp]
    \centering
    \includegraphics[width=0.9\hsize]{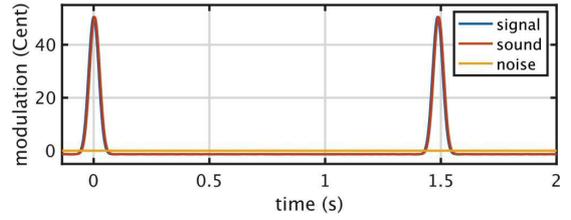}\\
    \vspace{-2mm}
    \caption{Recovered modulation and observed pulses processed from the electrically looped back signal and the acoustic signal through a headphone and a microphone.}
    \label{fig:recPulses}
\end{figure}
Figure~\ref{fig:recPulses} shows the analysis results of ``\textbf{System response analysis}'' in Fig.~\ref{fig:experimentSetting}.
The same $f_\mathrm{o}$ analysis procedure extracted the $f_\mathrm{o}$ values of the test signal (the electrically looped back signal) and the acquired acoustic signal (an omnidirectional condenser microphone acquired the sound produced by a noise-canceling headphone).
Filtering using time-reversed unit-TSPs and cross-correlation canceling procedure recovers the perturbation pulse and responses from the extracted $f_\mathrm{o}$ trajectories.
Please refer to the reference\cite{kawahara2020simultaneous} for details of these recovery procedures.
Note that the recovered pulse shapes are effectively the same (the electric signal yielded 50.38 Cent and the acoustic signal yielded 50.32 Cent at each peak.).\footnote{Figure~\ref{fig:recPulses} shows a slight delay in the acoustic pulse. The noise-canceling process of the headphone introduced about a 6~ms delay.}

Figure~\ref{fig:recPulses} illustrates that the proposed procedure provides accurate measurement of the $f_\mathrm{o}$ response to the perturbation.
It is important to note that the perturbation does not modify the subject's auditory feedback through the natural sidetone.
It enables us to measure the involuntary response to the $f_\mathrm{o}$ perturbation without disrupting the natural auditory-to-speech chain\footnote{Depending on the presentation level, the test signal may induce the Lombard's effect\cite{brumm2011evolution}.}.

\section{Measurement tool and examples}
We conducted preliminary experiments using the GUI-based test tool shown in Fig.~\ref{fig:toolGUI}.
Because of the COVID-19 pandemic, the first author played the experimenter and the subject roles.
This may introduce biases.
Please use these example results as illustration materials for proof of concept of the proposed method.
We made the tool and some examples accessible to everyone for them to be able to replicate the tests described in this article.
Please refer to the associated multimedia files.

\subsection{GUI-based test tool}
\begin{figure}[tbp]
    \centering
    \includegraphics[width=0.85\hsize]{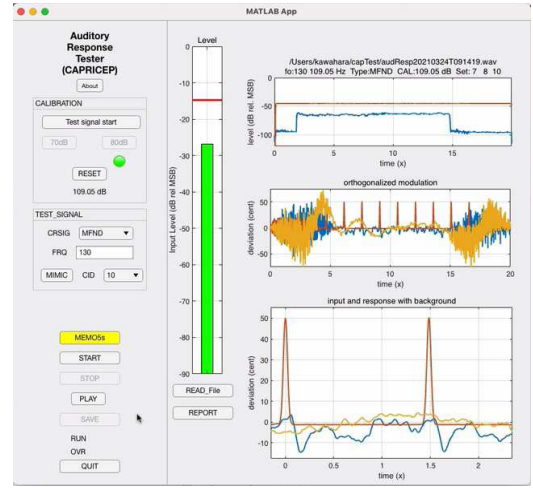}\\
    \vspace{-1mm}
    \caption{A snapshot of the GUI of the interactive test tool for measuring $f_\mathrm{o}$ response to the perturbation.}
    \label{fig:toolGUI}
\end{figure}
Figure~\ref{fig:toolGUI} shows a snapshot of the GUI of the interactive test tool for measuring $f_\mathrm{o}$ response to the perturbation.
This image shows a response to the missing fundamental (missing the first harmonic component) test signal.
The left part of the GUI is for operation, and the right part is for display the analysis result.
The center bar graph is an input level monitor.
The right three panels are the power level (top), 
recovered responses (middle), and the final response analysis result.
The title of the top panel shows the recorded file name and the test conditions.
Note that the analysis does not precede data saving.
Also, a separate log-file records all saving and analysis operations.
These are the built-in mechanism to prevent misconducts of experiments.

\subsection{Test conditions}
We used a miniature omnidirectional condenser microphone (Shure MX153T/O-TQG),
a noise-canceling circumaural headphone (SONY MDR-1RNC), an audio interface (ROLAND Rubix24), and a powered loudspeaker (IK Multimedia iLoud Micro Monitor).
The microphone and the R-channel audio interface output are connected R and L channels of the inputs of the audio interface.
We used a notebook computer (Apple MacBookPro 13 inches with 2.7GHz Intel Core i7 and 16~GB memory) for running the tool.
The placement of the microphone adopted the recommendation\cite{Rita2018ajsp}.
The sampling rate and the resolution was 44100~Hz and 24~bit.
The target $f_\mathrm{o}$ was 130~Hz, a comfortable pitch for the subject.

\subsection{Test procedure}
Before start experiments, the experimenter calibrates the acoustic input using pink noise and the calibration panel of the GUI.
A test session starts by clicking the ``\textbf{START}'' button.
In the beginning several seconds, the subject listens to the test sound to determine the target pitch of voicing. 
Then, the subject starts the sustained vowel keeping the pitch constant.
The test signal lasts in twenty seconds.
By clicking the ``\textbf{SAVE}'' button, it saves the test signal and the recorded voice, then response analysis starts and displays the results.
The attached media files consist of a movie showing an example test procedure.

\subsection{Example of analysis results}
\begin{figure}[tbp]
    \centering
    \includegraphics[width=0.9\hsize]{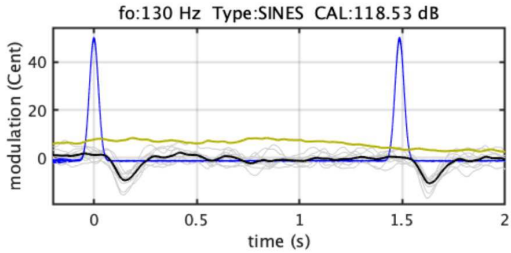}\\
    \vspace{-3mm}
    \caption{Response example to complex sound with 20 harmonic components.}
    \label{fig:complexSndEx}
    \vspace{-2mm}
\end{figure}
Figure~\ref{fig:complexSndEx} summarizes results of eight sessions using a test signal consisting of twenty harmonic components.
Each session used the test signal made from a different combination of unit-TSPs.
The different combination makes the non-linear time-invariant response independent each other.
The thin and light-gray solid lines show the obtained eight responses.
The thick black line shows the average response.
The blue line shows the frequency modulation pulse of the test signal.
The dark yellow line shows the averaged random responses.
Note that the average response is compensatory to the perturbation with a latency of around 100~ms.
Test signal having only the fundamental component prevented the subject from monitoring the pitch of the produced voice.
This condition made the $f_\mathrm{o}$ randomly drift away from the target value.
This behavior suggests that when using the test signal with harmonic components, the subject's auditory-to-speech chain of pitch regulation operates intact.
Therefore it is safe to state that the averaged response represents the involuntary response to the perturbation of this speech-to-speech chain.
These results illustrate the feasibility of the proposed method.

Each session lasts about one minute.
We tested under various conditions using different types of test signals.
The next section discusses the future possibilities we found from those test sessions.

\section{Discussion and future applications}
\begin{figure}[tbp]
    \centering
    \includegraphics[width=0.9\hsize]{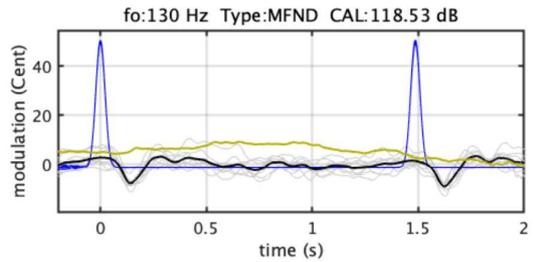}\\
    \vspace{-3mm}
    \caption{Response example to missing fundamental sound consisting of resolved harmonics from 2nd to 20th components}
    \label{fig:mfndEx}
    \vspace{-2mm}
\end{figure}

We conducted tests using a sum of harmonic components other than the fundamental component.
It is a missing fundamental signal.
Figure~\ref{fig:mfndEx} shows the summary of the results.
Note that the average of the response is close to that of test signals having the fundamental component (Fig.~\ref{fig:complexSndEx}).
This suggests that we can conduct this acoustic-to-speech chain experiment using a loudspeaker instead of using a headphone.
This is very useful.

We also tested using a sum of harmonic components from the eighth to the twentieth.
The average response to this test signal shows a very small compensatory behavior.
This difference in response may reflect the difference in the pitch salience of the test signals.
We speculate that the proposed procedure provides direct access to our internal pitch representation.
The immediate is to design test signals using several phase relations between harmonic components, such as cosine, alternating\cite{patterson1987jasa}, random, and Schroeder phase\cite{schroeder1970synthesis}.
These may lead to critical tests for testing various pitch perception models\cite{cheveigne2005pitch}.
The proposed method provides a non-invasive and quantitative assessment of auditory functions.




\section{Conclusions}
We designed a set of test signals using the mixture to modulate the fundamental frequency of artificial signals for testing the auditory-to-speech chain of pitch regulation. For testing the response, the experimenter presents the modulated signal aurally while the subject is voicing sustained vowels. We developed a tool for conducting this test quickly and interactively. We make it available as an open-source and also provide compiled GUI-based applications executable without requiring the MATLAB license. Preliminary tests using the tool revealed that the proposed method consistently provides compensatory responses with about 100~ms latency, representing involuntary control. Finally, we discuss future applications of the proposed method for objective and non-invasive auditory response measurements.

\section{Acknowledgments}
This work was supported by JSPS (Japan Society for the Promotion of Science) Grants-in-Aid for Scientific Research Grant Numbers JP18K00147, JP18K10708, and JP19K21618.

\bibliographystyle{IEEEtran}

\bibliography{kawaharaAudTbySp}


\end{document}